\documentclass[aps,twocolumn,superscriptaddress,floatfix]{revtex4-1} %linenumbers

\usepackage{lipsum}
\usepackage{graphicx}
\usepackage{amsmath,amssymb}
\usepackage{braket}
\usepackage{mathtools}
\usepackage{hyperref}

\usepackage{color}
\usepackage[normalem]{ulem}

\newcommand{\Z}{\mathbb{Z}}

\begin{document}
\title{Topologically-Protected Long Edge Coherence Times in Symmetry-Broken Phases}

\author{Daniel E. Parker}
\email[]{daniel\_parker@berkeley.edu}
\affiliation{Department of Physics, University of California, Berkeley, CA 94720, USA}

\author{Romain Vasseur}
\email[]{rvasseur@umass.edu}
\affiliation{Department of Physics, University of Massachusetts, Amherst, MA 01003, USA}

\author{Thomas Scaffidi}
\email[]{thomas.scaffidi@berkeley.edu}
\affiliation{Department of Physics, University of California, Berkeley, CA 94720, USA}

\newcommand{\zz}{\Z_2^{(\sigma)} \times \Z_2^{(\tau)}}

\date{\today}

\begin{abstract}

	We argue that symmetry-broken phases proximate in phase space to symmetry-protected topological phases can exhibit dynamical signatures of topological physics. This dynamical, symmetry-protected ``topological'' regime is characterized by anomalously long edge coherence times due to the topological decoration of quasiparticle excitations, even if the underlying zero-temperature ground state is in a non-topological, symmetry-broken state. The dramatic enhancement of coherence can even persist at infinite temperature due to prethermalization. We find exponentially long edge coherence times that are stable to symmetry-preserving perturbations, and not the result of integrability. 

\end{abstract}
\maketitle
 
Practical quantum computation requires systems with long coherence times. This has driven recent theoretical interest in the limits and causes of decoherence in quantum many-body systems where, typically, local quantum information is rapidly scrambled. One tactic to store and process quantum information is to use topological edge modes. Combining these with many-body localization~\cite{Mirlin05,Basko06,oganesyan2007localization,PalHuse,DavidRahulReview,EhudRonenReview,JoelRomainReview,2017arXiv171103145A,AbaninRev}, information can be protected for infinite times, even at effectively infinite temperature~\cite{PhysRevB.88.014206,1742-5468-2013-09-P09005,bahri2015localization,PhysRevB.89.144201,2015arXiv150806995Y}. Another avenue is to take advantage of prethermalization, wherein some observables retain memory of the initial state on a ``prethermal plateau'' before finally reaching their equilibrium values, leading to exponentially long coherence times~\cite{1742-5468-2012-11-P11020,1751-8121-49-30-30LT01,PhysRevB.90.165106,fendley2016strong,kemp2017long,else2017prethermal}.

\begin{figure}
	\centering
	\includegraphics[width=\linewidth]{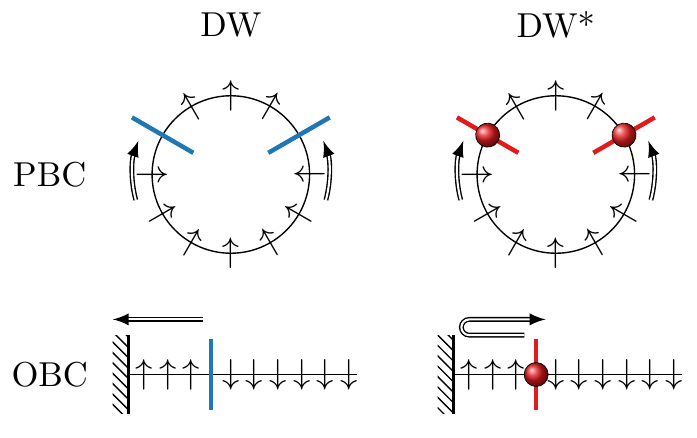}
	\caption{
		Sketch of the dominant processes that tunnel between the two ferromagnetic ground states. Domain walls (DW) are represented by blue bars, and their decorated counterparts (DW*) are red and carry a $\Z_2$ charge. Under periodic boundary conditions (PBC), the two types of domain walls are equivalent. With open boundary conditions (OBC), however, the decorated domain walls cannot be annihilated at the edges without breaking the symmetry, so will ``bounce off'' instead. Decorated domain walls are therefore unable to flip the edge spin without breaking the symmetry.
	}	
	\label{fig:cartoon}
\end{figure}

In this Letter we demonstrate an anomalous dynamical regime---characterized by long edge coherence times---that appears only in symmetry-broken phases proximate in phase space to symmetry-protected topological phases (SPTs)~\cite{doi:10.1146/annurev-conmatphys-031214-014740,PhysRevB.80.155131,PhysRevB.84.235128,PhysRevB.83.075102,PhysRevB.83.075103,Chen2011b,Pollmann2012,YuanMing2012,Levin2012,Chen1604,Chen2011}. The essential observation is that the presence of a nearby SPT phase can modify the nature of quasiparticle excitations even when the symmetry protecting the topological order is spontaneously broken at zero temperature. The topologically ``decorated''\cite{chen2014symmetry} quasiparticles inherited from the SPT cannot be created or annihilated at the edges of the system, leading to exponential increases in coherence times (see Fig. \ref{fig:cartoon}). Neither fine-tuning nor integrability are required. Even more remarkably, this protection of edge coherence remains at finite temperature and can persist all the way to infinite temperature thanks to prethermalization. Aspects of SPT physics, therefore, are retained in the dynamics even if the underlying zero-temperature ground state is symmetry-broken.

Though we will focus on SPTs, a motivation for this work comes from the ongoing experimental search for quantum spin liquids~\cite{0034-4885-80-1-016502,RevModPhys.89.025003,2018arXiv180402037K}, which are another form of topological paramagnets. Given the fact that many spin liquid candidate materials exhibit magnetically ordered ground states, the question arose as to whether remnants of a nearby topological paramagnetic phase could be detected in their dynamical properties. Indeed, such a ``proximate spin liquid'' regime was recently reported in $\alpha$-RuCl$_3$ \cite{Banerjee:2016aa,Banerjee1055}. In this Letter we answer this question in the affirmative, by providing an example of a proximate SPT regime whose anomalous dynamical properties are sharply defined. 

Below we define a simple model of a proximate SPT regime that demonstrates exponential enhancement in edge coherence times. To understand its dynamics, we consider the regular and decorated quasiparticles inherent to the model. This quasiparticle picture is confirmed at zero temperature, where we accurately predict the coherence times via perturbation theory. We then proceed to show that the regime is robust to symmetry-preserving perturbations, independent of integrability, and holds at all temperatures.

\begin{figure}
	\centering
	\includegraphics{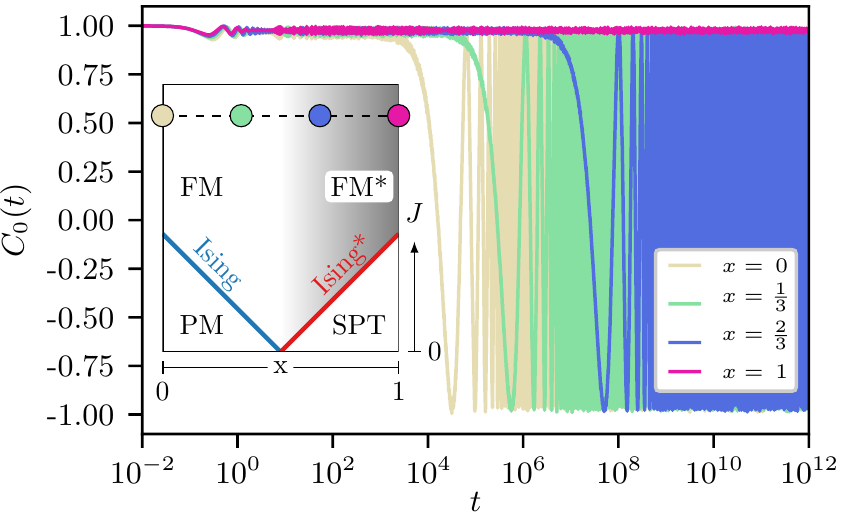}
	\caption{Autocorrelation of the edge spin at zero temperature computed with exact diagonalization (ED) for $14$ spins and OBC. Non-zero parameters are $(J,B) = (5.2,1.274)$.  \textit{Inset:} Sketch of the phase diagram for Eq. (1) as a function of $x$ and $J$. Phases are described in the text. The location of the dots corresponds to the data by color.}
\label{fig:T_0_results}
\end{figure}

\textit{Model and phase diagram}.
We rely on the simplest model of an SPT phase in one dimension, a variant of the Haldane chain~\cite{PhysRevLett.61.1029} protected by a global $\Z_2 \times \Z_2$ symmetry \cite{Chen1604,chen2014symmetry,PhysRevB.96.165124}. 
Consider a spin-$\tfrac{1}{2}$ chain with two alternating species, $\sigma$ and $\tau$, with a global $\Z_2^{\sigma} \times \Z_2^{\tau}$ symmetry generated by $\prod_i \sigma^x_i$ and $\prod_i \tau^x_i$. (We use the convention $\sigma_0, \tau_0, \sigma_1, \tau_1,\dots,\tau_{(L/2)-1}$ to label the $L$ spins.)  
We adopt a Hamiltonian
\begin{align}
	\hat{H}(x) \ &=J \hat{H}_{\text{FM},\sigma} + (1-x) \hat{H}_\text{PM} + x \hat{H}_\text{SPT} ,
%	H_\text{PM} \ &=\ - \sum_i \sigma_i^x  + B \tau_i^x \\
%	H_\text{SPT} \ &=\ - \sum_i \tau_{i-1}^z \sigma_i^x \tau_{i}^z
%		+ B \sigma_{i}^z \tau_i^x \sigma_{i+1}^z
		\label{eq:model}
\end{align}
where
$\hat{H}_{\text{FM},\sigma}=-\sum_i \sigma^z_i \sigma^z_{i+1}$,
$\hat{H}_\text{PM} \ =\ - \sum_i \sigma_i^x  + B \tau_i^x $,
$\hat{H}_\text{SPT} \ =\ - \sum_i \tau_{i-1}^z \sigma_i^x \tau_{i}^z
+ B \sigma_{i}^z \tau_i^x \sigma_{i+1}^z$, and $0 \le x \le 1$.
As shown in the inset of Fig.~\ref{fig:T_0_results}, this model interpolates between three different phases: a ferromagnet for the $\sigma$ spins at large $J$, a trivial paramagnet at small $J$ and $x$ near $0$, and an SPT (``topological paramagnet'') at small $J$ and $x$ near $1$.
Starting from either paramagnetic phase, $J$ drives an Ising transition to a ferromagnet for the $\sigma$ spins, and 
$B$ controls the energy scale for the $\tau$ spins, which remain paramagnetic across the whole phase diagram \footnote{Except at the topological phase transition $(x=1/2, J=0)$ where they become critical.}. This is the simplest version of the model; below we employ the generalization $\hat{H}(x) + \hat{V}$ where $\hat{V}$ includes generic symmetry-preserving perturbations to break integrability~\cite{SupMat}.

A standard result is that the two paramagnetic phases have the same bulk properties, but are different at the boundary: the SPT has a free spin-$\tfrac{1}{2}$ at each edge, which is protected as long as the $\Z_2 \times \Z_2$ symmetry survives~\cite{chen2014symmetry,Chen1604}. A lesser-known result is that these edge modes actually survive at the phase transition, leading to a ``topological'' variant of the Ising transition on the topological side (the red Ising* line), by forcing an anomalous conformal boundary condition~\cite{PhysRevX.7.041048,PhysRevB.97.165114,verresen2017topology}.
In the ferromagnetic phase, however, one would naively expect the topological physics to be lost since the protecting symmetry is spontaneously broken. 

\textit{Decorated quasiparticle picture}.
We show instead that the dichotomy between $x=0$ and $x=1$ extends not only to the Ising transition line, but also to the entire ferromagnetic phase beyond it. This is governed by the properties of quasiparticles. As there is no phase transitions, the static, ground state properties remain the same across the entire $x$ range, yet the nature of the dominant quasi-particle excitations changes dramatically with $x$.

As usual for a ferromagnet, quasiparticle excitations are domain walls, separating domains of opposite magnetization (for the $\sigma$ spins).
What is unusual, however, is that there are two kinds of domain walls in this model: the regular domain walls, generated by $H_\text{PM}$, and the ``decorated'' domain walls, generated by $H_\text{SPT}$. The latter kind is decorated in the sense that it carries a charge for the $\Z_2^\tau$ symmetry \cite{chen2014symmetry}.

This decoration is inconsequential in the bulk, where domain walls are always created or annihilated in pairs---but it has a drastic effect at the edge of the system. Flipping an edge spin changes the number of domain walls by $\pm 1$, which leads to a change in the total $\Z_2^\tau$ charge sector whenever the domain wall is decorated. Such a process necessarily breaks the $\Z_2^\tau$ symmetry and is therefore disallowed. This means that decorated domain walls cannot flip an edge spin without breaking the symmetry, while regular domain walls can. Note that the PM (resp. SPT) phase corresponds to the condensation of regular (resp. decorated) domain walls.

These considerations are, of course, irrelevant for static properties of the FM ground states, which contain no domain walls.
On the other hand, dynamical properties are dominated by the dynamics of domain walls, and it hence makes a difference whether they are decorated or not.
SPT proximity effects are thus invisible in static bulk properties, but are revealed in dynamical properties of the edge.
The remainder of the text will therefore be devoted to the dynamical properties of the model.

Let us consider the autocorrelation of the edge spin at temperature $T$, $C_T(t) = \mathrm{Re} \left\langle \sigma_0^z(t) \sigma_0^z(0) \right\rangle_T$. 
Fig.~\ref{fig:T_0_results} and Fig. 4 (a) show $C_T(t)$ for various cases, and Fig. 3 shows the coherence time  as a function of $x$, defined as the typical decay time of $C_T(t)$ \footnote{Specifically, $\tau \equiv \inf\{ t > 0:C(t)=1/e\}$.}.
As seen in Fig. 3, for OBC, the edge coherence time is exponentially larger at $x=1$ than at $x=0$, while no such increase is observed in the case of periodic boundary conditions.
This dramatic increase in edge coherence is due to the dominance of decorated domain walls in the region close to $x=1$ (dubbed FM*).

\begin{figure}
	\centering
	\includegraphics{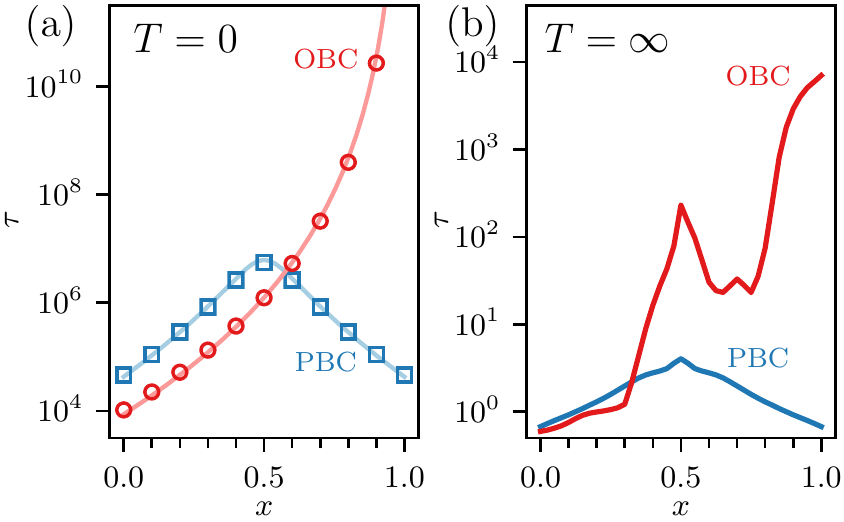}
	\caption{
		\textbf{($\mathbf{T=0}$)} Comparison of the coherence time (data) with its analytical prediction (lines)  Data is computed on $14$ spins via ED with parameters $(J, B) = (5.2,1.274)$. \textbf{($\mathbf{T = \infty}$)} Comparison of coherence times for OBC and PBC at infinite temperature on $14$ spins. The general trends are the same as at $T=0$. It was checked that the model is not integrable. (see Figs. \ref{fig:T_infinity_results} (c) and (d).)
}	
	\label{fig:T_0_x_predict_coherences}
\end{figure}

\textit{$T=0$ dynamics.} To confirm the quasiparticle picture we have outlined, we first work at zero temperature. Although the dynamics of a $T=0$ ferromagnet become trivial in the strict thermodynamic limit, we work at finite system sizes, which will provide a useful diagnostic of the ``hidden'' topological effects in the FM* region.
In this case, the notion of ``coherence time'' is nothing but the period of the Rabi oscillations between the two ground states, as seen in Fig.~\ref{fig:T_0_results}.  Deep in the ferromagnetic phase, there are indeed two nearly-degenerate ground states, $(\ket{\uparrow} \pm \ket{\downarrow})/\sqrt{2}$, where $\ket{\uparrow}$ (resp. $\ket{\downarrow}$) is a state with $\sigma_i^z=+1$ (resp. $-1$) and $\tau^x_i=+1$.
 The Rabi period is of course given by the inverse of the energy splitting $\Delta E$ between these two ground states. 
 While the coherence time $\tau$ is infinite in the thermodynamic limit for all $x$, one can see in Fig~\ref{fig:T_0_results} and~\ref{fig:T_0_x_predict_coherences} that its finite-size value has a systematic $x$ dependence --- it grows exponentially with $x$ --- thereby revealing a fundamental difference between the dynamics of the two sides.

  Within degenerate perturbation theory, the splitting $\Delta E$ is proportional to the tunneling rate from $\ket{\uparrow}$ to $\ket{\downarrow}$. 
  With PBC, the lowest order tunneling process occurs at order $L/2$ and corresponds to two domain walls being nucleated, propagating around the system, and annihilating each other. (See Fig. \ref{fig:cartoon}.) Such a process can occur for a pair of either regular or decorated domain walls, leading to 
\begin{equation}
	\Delta E_{\text{PBC}}(x) \propto \Delta E_{\text{DW}} + \Delta E_{\text{DW*}},
	\label{eq:PBC_splitting}
\end{equation}
 where $\Delta E_{\text{DW}} = \left( \frac{1-x}{4(J+xB)} \right)^{L/2}$ is the contribution for regular domain walls and $\Delta E_{\text{DW*}}(x) = \left( \frac{x}{4(J+(1-x)B)} \right)^{L/2}$ is the contribution for decorated domain walls.
Note that \eqref{eq:PBC_splitting} is symmetric under $x \leftrightarrow 1-x$, reflecting the equivalence of the two kinds of domain walls for PBC.

Open boundary conditions change the situation significantly.
Given the facts that (i) going from one ground state to another involves flipping all the $\sigma$ spins, including at the edges, and (ii)  decorated domain walls cannot flip an edge spin, it is clear that only regular domain walls contribute to the splitting. (See Fig. \ref{fig:cartoon} for illustration.) Hence
\begin{equation}
	\Delta E_{\text{OBC}}(x) \propto \Delta \widetilde{E}_{\text{DW}},
	\label{eq:OBC_splitting}
\end{equation}
where the tilde signifies that the regular domain wall contribution is slightly modified compared to PBC: $\Delta \widetilde{E}_{\text{DW}} = \frac{1}{1-x} \left( \frac{1-x}{2(J+xB)} \right)^{L/2}$.
This is manifestly {\textit{asymmetric}  under $x \leftrightarrow 1-x$ and indeed vanishes in the limit $x \to 1$, leading to a diverging coherence time on the topological side. Fig.~\ref{fig:T_0_x_predict_coherences} (a) shows that Eqs.~\eqref{eq:PBC_splitting} \&~\eqref{eq:OBC_splitting}} accurately predict the coherence times in this simple limit. We have checked that adding generic symmetry-preserving perturbations, including processes which can ``un-decorate'' a domain wall, will remove the divergence, but preserve the phenomena of exponentially longer coherence at $x=1$ than $x=0$~\cite{SupMat}.

\begin{figure*}
	\centering
	\includegraphics[width=\textwidth]{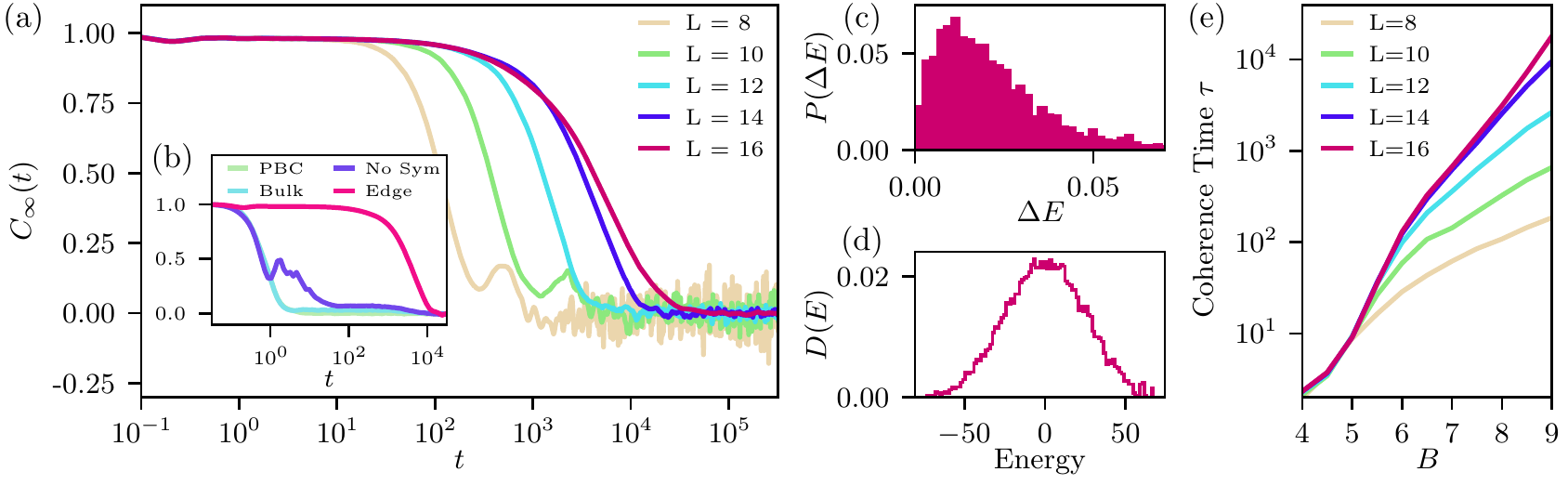}
	\caption{
	\textbf{(a)} Autocorrelation $C_\infty(t)$ at $x=1$ and $T=\infty$ under OBC and varying system size. $C_\infty(t)$ remains close to one for a time $\tau$ until it drops to its thermal value of $0$, and
		$\tau$ increases exponentially with system size until its saturation.
		\textbf{(b)} The same autocorrelation $C_\infty(t)$ under various conditions on $14$ sites. `Edge' is same as in the main panel, `bulk' corresponds to $\sigma_{L/4}^z$, `PBC' corresponds to periodic boundary conditions, and `No Sym' corresponds to a system where the $\Z_2 \times \Z_2$ symmetry was broken explicitly with edge perturbations $\sigma_0^x \tau_1^z$ and $\sigma_0^y \tau_1^z$.
		\textbf{(c)} Histogram of the differences in adjacent energy levels and \textbf{(d)} normalized density of states in the $\Z_2 \times \Z_2$ even/even sector on $16$ spins.
	\textbf{(e)} Dependence of the coherence time for $x=1$ on $B$, which sets the energy scale for the $\tau$ spins. One can already see the saturation with system size for smaller values of $B$. Numerical details are given in the Suppemental Material \cite{SupMat}.
	}
	\label{fig:T_infinity_results}
\end{figure*}

\textit{$T > 0$ Dynamics.} 
At non-zero temperatures, there is a finite density $\rho \sim{} e^{-\Delta/T}$ of domain wall quasiparticles, where $\Delta$ is the energy gap of the excitation~\cite{sachdev1997low,sachdev2011quantum}. For $x$ close to $1$, decorated domain walls have a lower gap than regular ones, and therefore are expected to dominate the dynamics at low $T$. For higher $T$, on the other hand, there is a finite density of both kinds of domain walls, so the naive expectation is that topological effects will disappear.

Surprisingly, we find instead that the enhancement of coherence from $x=0$ to $x=1$ with open boundary conditions persists even at $T = \infty$ (Fig. \ref{fig:T_0_x_predict_coherences} and Fig.~\ref{fig:T_infinity_results}).
(The results for intermediate temperatures $0 < T < \infty$ are similar and are described in the Supplemental Material~\cite{SupMat}.)
We have checked that this behavior does not rely on integrability. The level spacings, shown in Fig.~\ref{fig:T_infinity_results}.(c) have good level repulsion with a shape characteristic of GOE statistics~\cite{oganesyan2007localization}. The many-body density of states in panel (d) is  normally distributed, as is required to be representative of the thermodynamic limit~\cite{PAPIC2015714} (see~\cite{SupMat} for more details). As expected for a non-integrable system~\cite{kemp2017long,else2017prethermal}, while the coherence time initially increases exponentially with $L$, it eventually saturates to a $L$-independent value. This behavior can be seen in Fig. \ref{fig:T_infinity_results} (a) and (e).

To understand the survival of coherence at infinite temperature, we appeal to the physics of prethermalization. As shown in Fig. \ref{fig:T_infinity_results}.(e), the dominant parameter that controls the coherence time is $B$, which sets the energy scale for the $\tau$ spins.
It is therefore instructive to consider the case of $B \gg 1$ and to rewrite the Hamiltonian as
\begin{equation}
	\hat{H}= - B\left[ x \hat{N}^* + (1-x) \hat{N}\right] + \hat{V}_p,
\end{equation}
where $\hat{N}^*=\sum_i \sigma_{i}^z \tau_i^x \sigma_{i+1}^z$, $\hat{N}= \sum_i \tau^x_i$ and $\hat{V}_p$ contains all the $\mathcal{O}(1)$ terms that are independent of $B$. The operator $\hat{N}^*$ counts the number of ``mismatched decorations'': domain walls without a $\Z_2^\tau$ charge attached, or $\Z_2^\tau$ charges without a domain wall. 

While there are symmetry-respecting processes which can flip the edge spin, one can show that they necessarily have to change the $\hat{N}^*$ sector. (For instance, $\sigma_0^x$ anticommutes with $\hat{N}^*$.)
Such processes are exponentially suppressed due to the so-called ADHH theorem~\cite{abanin2017rigorous}. The theorem states, roughly, that if $e^{2\pi i \hat{N}^*} = 1$ and $\hat{N}^*$ is a sum of commuting projectors --- which is indeed the case here --- then $\hat{N}^*$ is approximately conserved until at least a (quasi)-exponentially long time $\tau \sim e^{B x/h}$, where $h$ is the norm of the second-largest term after $\hat{N}^*$. (See \cite{abanin2017rigorous} for the precise statement.) For $x$ close to $1$ \footnote{Roughly when $1-x < 1/B$}, the second largest term is in $\hat{V}_p$, so $h$ is $\mathcal{O}(1)$ and we expect $\tau \sim{} e^{Bx}$. We find indeed in Fig. \ref{fig:T_infinity_results} (e) that the large-$L$ saturation value of $\tau$ increases exponentially with $B$ for $x=1$. For $x$ away from $1$, the second largest term is $\hat{N}$, leading to $\tau \sim e^{x/(1-x)}$ (excluding special values of $x$ at which the sum of $N$ and $N^*$ have integer spectrum, leading to extra peaks in the coherence, see Fig \ref{fig:T_0_x_predict_coherences} (b) ).

This enhancement of the coherence is ``topological'', since only the coherence of the edge is exponentially enhanced and, unlike previous applications of the ADHH theorem~\cite{kemp2017long,else2017prethermal}, it is also symmetry-protected. Explicitly, this means that adding terms which break the $\Z_2 \times \Z_2$ symmetry can immediately destroy the anomalously long edge coherence times. The term $\sigma_0^x \tau_1^z$, for instance, commutes with $\hat{N}^*$ but breaks the $\Z_2^\tau$ symmetry and is able to flip the edge spin and suppress the coherence, as  shown in Fig. \ref{fig:T_0_x_predict_coherences} (b).  This provides a clear example of (prethermal) SPT physics even at infinite temperature, in a regime where the protecting symmetry is spontaneously broken at zero temperature.

\textit{Discussion}. 
	We have demonstrated the existence of a proximate SPT regime, characterized by anomalously long edge coherence times. The key to the model's dynamics is the behavior of its two species of quasiparticles: regular and decorated domain walls. The decorated domain walls, which are inherited from the SPT phase, cannot be created or annihilated near the edges of the system without breaking the symmetry, giving rise to a dramatic increase in edge coherence.
	 In the special case of zero temperature, we confirmed the quasiparticle picture within perturbation theory. We have shown that the phenomena is robust; the enhancement of edge coherences is stable to symmetry-preserving perturbations, integrability-breaking perturbations and, via prethermalization, survives at all temperatures.

	The existence of a proximate SPT regime has several broader implication. It shows how to advantageously combine two different ways to reach long coherence times which were, up to now, thought to be antinomic: symmetry-protected topological effects and ferromagnetism. 
	Furthermore, we have shown that topological effects are not strictly confined by their phase boundaries, but can infect the finite temperature dynamics of nearby quantum phases. We expect this to extend beyond SPTs to spin liquids and other topological phases. 

\textit{Acknowledgments}. 
	We thank Maksym Serbyn, Ehud Altman and Joel Moore for insightful discussions. We acknowledge support from NSF Graduate Research Fellowship Program NSF DGE 1752814 (D.P.), the Emergent Phenomena in Quantum Systems initiative of the Gordon and Betty Moore Foundation (T.S.), and University of Massachusetts start-up funds (R.V.).

	\bigskip

	\bibliography{references}

%%%%%%%%%% Merge with supplemental materials %%%%%%%%%%
\pagebreak
\begin{widetext}
\begin{center}
	\makeatletter
\textbf{\large Supplemental Materials: \@title}
\makeatother
\end{center}
\end{widetext}

%%%%%%%%%% Merge with supplemental materials %%%%%%%%%%
%%%%%%%%%% Prefix a "S" to all equations, figures, tables and reset the counter %%%%%%%%%%
\setcounter{equation}{0}
\setcounter{figure}{0}
\setcounter{table}{0}
\setcounter{page}{1}
\makeatletter
\renewcommand{\theequation}{S\arabic{equation}}
\renewcommand{\thefigure}{S\arabic{figure}}
\renewcommand{\bibnumfmt}[1]{[S#1]}
\renewcommand{\citenumfont}[1]{S#1}
%%%%%%%%%% Prefix a "S" to all equations, figures, tables and reset the counter %%%%%%%%%%

This Supplemental Material provides additional technical details. We first define the perturbations that are added to the Hamiltonian, and some details of the numerics.  We then discuss the model at intermediate temperatures, $0 < T < \infty$, and show there is a smooth cross-over. Finally, we comment on the spectral statistics of our model.

%In this Appendix we give some additional technical details. We first define the perturbations added to break integrability. We then discuss the intermediate temperature dependence of the model, showing it is in line with the predictions in the main text. Finally we comment on the spectral statistics of our model and the effects of finite system size.

\section{Perturbations and Numerics}

Recall from the main text that we adopt a Hamiltonian $\hat{H}(x) + \hat{V}$ where $\hat{H}$ is composed of competing paramagnetic, SPT, and ferromagnetic terms. While $\hat{H}(x)$ alone includes all the essential physics, it is unnecessarily fine-tuned; our results do not depend on the specific form of the Hamiltonian and are unchanged when generic perturbations $\hat{V}$ are added.

Let us describe carefully what we mean by generic. As mentioned in the main text, the enhanced coherence time is symmetry-protected, so $\hat{V}$ must obey the $\Z_2 \times \Z_2$ symmetry. Furthermore, with periodic boundary conditions, the spectra of $\hat{H}(x)$ and $\hat{H}(1-x)$ are identical, since without boundaries the SPT physics should be inconsequential, and this should continue to hold in the presence of perturbations. Generic perturbations are therefore ones that both respect $\Z_2 \times \Z_2$ and preserve the $x \leftrightarrow 1-x$ correspondence.

We first consider $x=0$. Note that $\hat{H}(x=0)$ consists of the standard Ising model for the $\sigma$ spins and a paramagnet for the $\tau$'s. To this we add the perturbation
\begin{equation}
	\begin{aligned}
	\hat{V}_0 &= - \sum_i g_1 \tau_i^z \tau_{i+1}^z + g_2 \sigma_i^x \sigma_{i+1}^x + g_3 \tau_i^x \tau_{i+1}^x\\
	\ &\hspace{4em} + g_4 \sigma_i^x \tau_i^x + g_5 \sigma_i^z \tau_i^z \sigma_{i+1}^z \tau_{i+1}^z.
\end{aligned}
	\label{eq:perturbations}
\end{equation}
Here the $g_1$ term gives a finite correlation length to the $\tau$ spins; the $g_2$ and $g_3$ terms take the $\sigma$ and $\tau$ spins away from integrability; and finally the $g_4$ and $g_5$ terms couple the $\sigma$'s and $\tau$'s together. One can check these are the simplest local perturbations (i.e., involving the smallest number of spins) compatible with the $\Z_2 \times \Z_2$ symmetry. We consider a regime of parameters such that the $\tau$'s are always paramagnetic, while the $\sigma$'s form a ferromagnet at zero temperature. 

We now extend these perturbations to all $x$ by employing a tool from the field of SPT's. Define the ``decorating'' local unitary operator $\hat{U}$, whose action is to exchange regular and decorated domain walls, via $\sigma_i^x \to \tau_{i-1}^z \sigma_i^x \tau_{i}^z$, $\sigma_i^y \to \tau_{i-1}^z \sigma_i^y \tau_i^z$, $\sigma_i^z \to \sigma_i^z$, and similarly for the $\tau$ spins. Under periodic boundary conditions, this commutes with the $\Z_2 \times \Z_2$ symmetry and satisfies $\hat{U}^\dagger H(x) \hat{U} = H(1-x)$ . We therefore define the perturbations for all $x \in [0,1]$ by
\begin{align}
	\hat{V} = x \hat{V}_0 + (1-x) \hat{U}^\dagger \hat{V}_0 \hat{U}.
	\label{eq:perturbations_2}
\end{align}
Note with open boundary conditions there are some terms such as $\tau_{-1}^z \tau_0^z$ which do not make sense. Following the standard procedure for SPT's, these  are omitted. To ensure the dynamics are generic at the edge, we add terms $g_e \sigma_0^x + B g_e \tau_{(L/2)-1}^x$ for some $O(1)$ parameter $g_e$.

Equation \eqref{eq:perturbations_2} constitutes generic perturbations to $\hat{H}(x)$ which preserve the $\Z_2 \times \Z_2$ symmetry and the correspondence $x \leftrightarrow 1-x$ under periodic boundary conditions. Our standard choice of parameters, used in Fig. 3 and 4 in the main text, is $y = 1.5766$, $B = 8.4238$, $g_1 = 3.2654$, $g_2 = -0.1872$, $g3=0.1121$, $g_4 = 0.3518$, and $g_5 =0.2804$. These values were chosen randomly and it was checked that they avoid resonances (see below).

\begin{figure}
	\centering
	\includegraphics{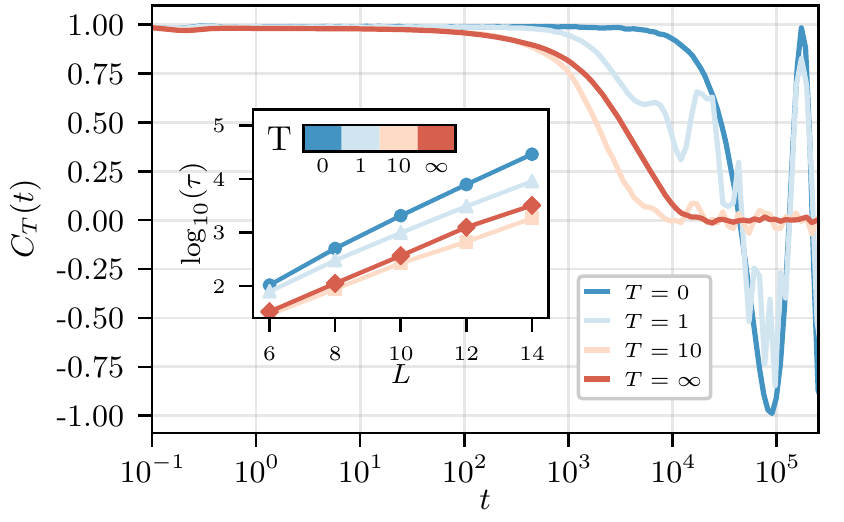}
	\caption{The autocorrelation $C_T(t)$ at four different temperatures as a function of time and $x=1$. \textit{Inset}: Coherence time as a function of the number of sites at four temperatures. Hamiltonian parameters are the ``standard'' ones given in the text.}
	\label{fig:temp_dependence}
\end{figure}

\section{Temperature Dependence}

In the main text it was discussed how the behavior of the edge spin crosses over from Rabi oscillations at zero temperature to prethermal behavior at infinite temperature. Fig. \ref{fig:temp_dependence} below shows the autocorrelation $C_T(t)$ for the edge spin at four different temperatures. For all curves, there is an initial short-time transient until about $t \approx 10^{-1/2}$, after which the value of the autocorrelation is roughly constant at $C_T(t) \approx 0.99$ until $t \approx 10^3$. At this point, the behaviors differ by temperature. At $T=0$, $C(t)$ undergoes an oscillation between $-1$ and $1$ with period of $t \approx 0.8 \times 10^5$, governed by the process of tunneling between ground states discussed in the main text. For $T =1$, there are thermal populations of quasiparticles, allowing additional processes to contribute. This gives further oscillations on top of the zero temperature one. By $T=10$, so many processes contribute, all with different periods and phases, that destructive interference occurs and $C_T(t)$ becomes approximately $0$ for $t > 10^4$. This is the same at $T=\infty$, where one can clearly see the prethermal behavior until the coherence dips down to zero. The small oscillations after $t \approx 10^4$ are non-universal. Examining Fig. 4 of the main text, one can see these decrease with system size. The inset shows the growth of the coherence time with system size. This is, however, a finite-size effect and the coherence eventually reaches saturation.

\section{Spectral Statistics and Finite-Size Effects}

There are several effects which can increase coherence times but are the result of fine-tuning rather than a generic phenomena. We show that the most common---integrability, finite-size bands, and accidental resonances---do not apply here.

Quantum integrable systems, which enjoy an extensive number of conserved quantities, can often preserve coherences for long times \cite{1751-8121-49-30-30LT01,kemp2017long}.
However, integrability is quite ``fragile'', and is removed by infinitesimal perturbations in the thermodynamic limit. In Fig 2. of the main text, we have seen that adding the perturbation $\hat{V}$ reduces the coherence time relative to the unperturbed case, but does not change the overall pattern of much longer coherence for $x=1$ than $x=0$. This already suggests our coherence times are not due to integrability. To check this more stringently, we examine the spectral statistics. Working in the $\Z_2 \times \Z_2$ even/even sector of the Hamiltonian, the full spectrum was computed on $16$ sites. For an integrable system, eigenvalues adjacent in energy usually lie in different sectors of the conserved quantities, so their differences $\delta_n = E_{n+1} - E_n$ are uncorrelated. This leads to a distribution of $\delta_n$'s peaked at $0$, whereas for a non-integrable system, the differences of levels approach that of the generalized orthogonal ensemble of random matrices, otherwise known as the Wigner Surmise.

\begin{figure}
	\centering
	\includegraphics{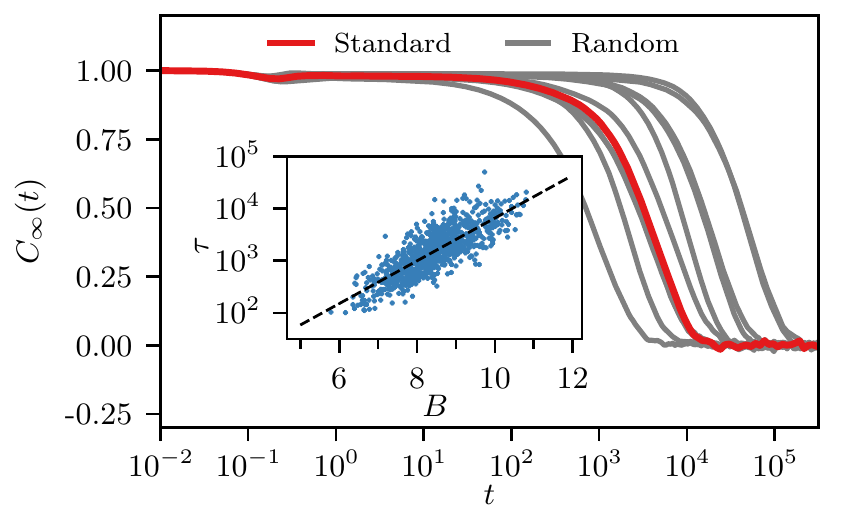}
	\caption{Autocorrelation on 14 spins, $x=1$, and $T=\infty$ for the standard and randomized choice of Hamiltonian parameters, as described in the text. One can see that the random autocorrelations are distributed around the standard one. \textit{Inset:} coherence times on 12 spins at $x=1$ and $T=\infty$ versus $B$, with a linear regression. One can see that the parameter $B$ accounts for most of the variation in coherence time.}
	\label{fig:random_parameters}
\end{figure}

Panel (c) of Fig. 3 in the main text shows the differences are indeed distributed following the Wigner Surmise, as expected. To assess this more quantitatively, we computed the $R$-statistic~\cite{oganesyan2007localization}, defined as the ratio of adjacent differences in eigenvalues:
\begin{align}
	r_n = \frac{\min\{\delta_n, \delta_{n-1}\}}{\max\{\delta_n,\delta_{n-1}\}}.
	\label{eq:R_statistic}
\end{align}
For an integrable system, the expectation over all eigenvalues is $\braket{r} \approx 0.385$, while for a non-integrable system, $\braket{r} \approx 0.528$ in the thermodynamic limit~\cite{oganesyan2007localization}. We measured $\braket{r_n} \approx 0.49$ with the ``standard'' parameters, showing our model is far from integrability. (For the other sectors, $\braket{r_n}$ is almost exactly the same.) Indeed, this is no surprise since the perturbation $\hat{V}$ was designed to remove integrability from both the $\sigma$ and $\tau$ spins. 

Another aspect to consider is the distribution of the eigenvalues $E_n$, i.e. the density of states. For a generic, non-integrable system, we expect them to be normally distributed, with no features. (See, e.g. \cite{PAPIC2015714} for a discussion.) In systems which are too small to capture the thermodynamic limit, or have some integrable structure, there are often bands or other quasiperiodic features visible. These features are frequently associated with increased coherences times because, to high order in perturbation theory, a quasiparticle can be ``trapped'' in energy in the band it starts in, limiting the number of final states it can scatter to. At larger system sizes, the number of eigenvalues in each band grows, as does their width, until they touch and hybridize, forming a smooth density of states. Panel (d) of Fig. 3 in the main text shows the density of states is featureless and approximately Gaussian distributed. The system sizes we use are therefore large enough to be considered generic from this perspective.

We have also taken pains to avoid accidental resonances. In higher-order perturbation theory, resonances can form between terms whose magnitudes have small number ratios (e.g, $0.6 \sigma_i^z \sigma_i^{z+1}$ and $0.8 \sigma_i^y$ with a ratio of $3$ to $4$). It is therefore generally recommended to take parameters free from such coincidences, such as irrational numbers with non-coinciding continued fraction expansions. Here we take a slightly different approach: we randomly vary the parameters and observe it has no effect on the physics. In Fig. \ref{fig:random_parameters} we compare the autocorrelation with ``standard'' parameters given above to randomly chosen parameters. Explicitly, we compute the autocorrelation where the parameters $J, B, g_1, g_2, g_3, g_4, g_5$ are randomized by multiplying the standard values, given above, by random numbers from the normal distribution with mean $1$ and standard deviation of $0.1$. Ten realizations of this randomization are shown in Fig. \ref{fig:random_parameters}. All the autocorrelations fall close to the standard one, and are distributed around it. The changes in coherence time are mainly caused by the variation in the parameters $B$, as is shown in the inset to Fig. \ref{fig:random_parameters}. This confirms that the standard parameters are suitable generic; accidental resonances are not the cause of the long coherence times observed here.

We also note that, although our numerics has focused on the left-most $\sigma$ spin, the autocorrelation is symmetric under inversion. In other words, the $\sigma$ spins at both sides have long coherence times, as one would expect in analogy with an SPT.

We may thus conclude that our model is not fine-tuned in any obvious way.

%\bibliography{references2}

\end{document}